\documentclass[%
reprint,
 amsmath,amssymb,
 aps,
]{revtex4-1}

\usepackage{graphicx}
\usepackage{dcolumn}
\usepackage{bm}

\begin{document}
\title{Crossing the dark matter soliton core: a possible reversed orbital precession}
\author{Man Ho Chan$^{\dag}$, Chak Man Lee}
\address{Department of Science and Environmental Studies, The Education University of Hong Kong, Tai Po, New Territories, Hong Kong, China
\\
Correspondence Email: $^{\dag}$chanmh@eduhk.hk\\
}
\begin{abstract}
The ultra-light dark matter (ULDM) model has become a popular dark matter scenario nowadays. The mass of the ULDM particles is extremely small so that they can exhibit wave properties in the central dark matter halo region. Numerical simulations show that a soliton core with an almost constant mass density would be formed inside the ULDM halo. If our Galactic Centre has a dark matter soliton core, some of the stars orbiting about the supermassive black hole (Sgr A*) would be crossing the soliton core boundary. In this article, we report the first theoretical study on how the dark matter soliton core near the Sgr A* could affect the surrounding stellar orbital precession. We show that some particular stellar orbital precession may become retrograde in direction, which is opposite to the prograde direction predicted by General Relativity. We anticipate that future orbital data of the stars S2, S12 and S4716 can provide crucial tests for the ULDM model for $m \sim 10^{-19}-10^{-17}$ eV.
\end{abstract}
 \maketitle

\section{Introduction}
In the past few decades, many direct-detection (e.g. XENON-1T) \cite{Amole,Aprile,Aalbers} and indirect-detection (e.g. radio and gamma-ray detection) \cite{Ackermann,Chan3,Chan4} experiments were deployed to search for the signal of weakly-interacting massive particle (WIMP) dark matter. However, no promising signal has been obtained so far \cite{Bertone2}. Therefore, some suggest another extreme that dark matter may be made of bosons with very small mass (i.e. the ultra-light dark matter, ULDM). This proposal can simultaneously solve some long-standing problems in particle physics \cite{Bertone,Marsh}. Since the mass of the ULDM is very small, they will exhibit wave nature at small scales. Interestingly, the wave behaviour of dark matter would undergo interference and it would finally form a soliton core at the central region of the ULDM halo. Numerical simulations show that the soliton core formed by the ULDM would be quite universal, which depends on the mass of the ULDM particle $m$ and the halo mass $M_h$ \cite{Schive,Safarzadeh}.

The ULDM model has become more popular nowadays because it can form dark matter cores observed in many dwarf galaxies (i.e. solving the core-cusp problem) \cite{Hu}. The ULDM can behave like cold dark matter on large scales while it can demonstrate core-like structures at small scales \cite{Hu}. Various studies have put constraints on the mass of the ULDM, such as using the data of galaxies \cite{Safarzadeh,Marsh2,Desjacques,Maleki,Li,Chan}, Lyman-$\alpha$ forest \cite{Rogers}, halo mass function \cite{Schutz}, supermassive black hole \cite{Stott,Monica,Saha}, and gravitational waves \cite{Hannuksela}. Nevertheless, many of the constraints are contradictory to each other and no robust conclusion has been made so far. Overall speaking, recent studies seem to favour $m \ge 10^{-20}$ eV \cite{Rogers,Hui} and some even suggest $m \ge 10^{-18}$ eV \cite{Desjacques}.

In this article, we report the first theoretical study on how the ULDM dark matter soliton core could affect the surrounding stellar orbital precession, which have not been discussed previously. We surprisingly discover that some stellar orbits at the deep Galactic Centre might possibly be reversal in the orbital precession if a dark matter soliton exists. This phenomenon depends on the position of the stellar orbits and the actual value of $m$. By using the information of the stellar cluster at the deep Galactic Centre (the S-Star cluster), we can test a wide range of ULDM mass ($m \sim 10^{-19}-10^{-17}$ eV), which is currently one of the popular ranges in theories (e.g. string axions) \cite{Marsh,Hui}. Future data from GRAVITY collaboration and the orbital data of the newly discovered stars (e.g. S4716 star) near the Sgr A* can provide crucial tests to verify or falsify the ULDM soliton model. This can greatly enhance our understanding of the nature of dark matter.

\section{Equation of motion for ULDM soliton core model }
For a Schwarzschild black hole, the spherical symmetric space-time metric is
\begin{equation}
ds^2=A(r)c^2dt^2-B(r)dr^2-r^2(d\theta^2+\sin^2\theta d\phi^2),
\label{metric}
\end{equation}
where $(r, \theta,\phi)$ are the spherical coordinates, $A(r)=1-r_s/r$ with $r_s=2GM_{\rm BH}/c^2$, and $B(r)=1/A(r)$. The equation of motion of a star orbiting about a supermassive black hole on the fixed plane at $\theta=\pi/2$ is given by:
\begin{equation}
\frac{d^2u}{d\phi^2}+u=\frac{GM_{\rm BH}}{L^2}+3\frac{GM_{\rm BH}}{c^2}u^2,
\label{trajectory}
\end{equation}
where $u=1/r$ and $L$ is the angular momentum. The solution of $u(\phi)$ is approximately given by
\begin{equation}
u(\phi) \approx \frac{GM_{\rm BH}}{L^2}[1+e\cos(\phi-\epsilon\phi)],
\end{equation}
where $e$ and $\epsilon$ are constant. Therefore, the precession angle of the stellar orbit per period due to the General Relativistic effect is
\begin{equation}
\Delta\phi_{\rm GR}=2\pi\epsilon \approx 6\pi \frac{G^2M_{\rm BH}^2}{c^2L^2} \approx 6\pi \frac{GM_{\rm BH}}{c^2a(1-e^2)}.
\label{precession}
\end{equation}

If the ULDM soliton core exists, the ULDM core mass would be much smaller than the supermassive black hole mass. Following the method of perturbation, we can re-write the equation of motion by adding a ULDM component as
\begin{equation}
\frac{d^2u}{d\phi^2}+u \approx \frac{GM(u)}{L^2}+3\frac{GM_{\rm BH}}{c^2}u^2,
\label{trajectory1}
\end{equation}
where $M(u)=M_{\rm BH}+M_{\rm DM}(u)$ and $M_{\rm DM}(r)$ is the enclosed mass of the ULDM halo. Note that the general relativistic term $3GM_{\rm BH}u^2/c^2$ can be regarded as a first order correction term of the Newtonian equation of motion. Since we have $M_{\rm DM}(u) \ll M_{\rm BH}$, adding a small term representing the effect of the ULDM component in the general relativistic term would further generate a second order correction term for the equation of motion, which is negligible. Therefore, there is no change in the general relativistic term in Eq.~(5) compared with that in Eq.~(2).

In our present numerical simulations, the central ULDM density profile is given by \cite{Schive,Safarzadeh}
\begin{equation}
\rho_{\rm DM}(r)=\frac{\rho_0}{(1+Ar^2)^8},
\end{equation}
where $A=9.1\times 10^{-2}r_c^{-2}$,
\begin{equation}
\rho_0=1.9 \left(\frac{m}{10^{-23}~{\rm eV}} \right)^{-2} \left(\frac{r_c}{\rm kpc} \right)^{-4}~M_{\odot}~{\rm pc^{-3}},
\end{equation}
and $r_c$ is the soliton core radius, which is given by
\begin{equation}
r_c=16 \left(\frac{m}{10^{-23}~\rm eV} \right)^{-1} \left(\frac{M_h}{10^9M_{\odot}}\right)^{-1/3}~{\rm kpc},
\end{equation}
with $M_h$ being the total halo mass. Therefore, the enclosed mass of the ULDM halo is
\begin{equation}
M_{\rm DM}(r)=\int_0^r 4 \pi r'^2 \rho_{\rm DM}(r')dr'.
\end{equation}
The final stellar orbit $r(\phi)=1/u(\phi)$ could be obtained by solving Eq.~(\ref{trajectory1}) numerically. Here, the central ULDM density profile is indeed embedded in a galactic regular dark matter density profile (e.g. the Navarro-Frenk-White density profile). For $r \gg r_c$, the ULDM density would follow the galactic regular dark matter density profile \cite{Schive}. 

\section{Results}
We first use the S2 star as an example to demonstrate the effect on the orbital precession angle due to the ULDM model. The motion of the S2 star has been monitored for almost three decades \cite{Eckart,Ghez98,Schodel,Eisenhauer,Ghez03,Ghez08,Heibel}. The S2 star has finished at least one period (orbital period $\approx 16$ years) since our first observation on it. Therefore, we have already obtained rich information about the S2 star and the constraints on the mass of the supermassive black hole.

If a ULDM soliton core exists in our Galactic Centre, numerical simulations can predict the size of the core radius $r_c$ and the central core density $\rho_c$ \cite{Schive,Safarzadeh}. The core radius and the central core density depend on both $m$ and the total halo mass $M_h$ \cite{Schive,Safarzadeh}. Recent observations of the Gaia satellite have constrained the total halo mass to be $M_h=1.08^{+0.20}_{-0.14} \times 10^{12}M_{\odot}$ \cite{Cautun}. This is consistent with the results obtained from other studies, such as $M_h=1.17^{+0.21}_{-0.15} \times 10^{12}M_{\odot}$ in \cite{Callingham} and $M_h=(1.16 \pm 0.24) \times 10^{12}M_{\odot}$ in \cite{Deason}. In the following analysis, we will adopt a wider possible range of the halo mass $M_h=(1.0-1.4) \times 10^{12}M_{\odot}$. Therefore, the central core density can be written in terms of $m$ only.

Recent observations have precisely determined the orbital parameters of the S2 star, including the semimajor axis $a$, eccentricity $e$, and the mass of the supermassive black hole $M_{BH}=(4.154 \pm 0.014) \times 10^6M_{\odot}$ \cite{Abuter}. By adding the ULDM component to the equation of motion, we can theoretically simulate the subsequent orbit of the S2 star for different $m$. We plot the predicted precession angle for the S2 star against $m$ in Fig.~1. We can see that for $m \sim 10^{-18}-10^{-17}$ eV, the effect due to the ULDM is large enough to change the orbital precession from prograde (positive precession angle) to retrograde (negative precession angle). However, recent observations have constrained the precession angle to be $\Delta \phi=(0.22 \pm 0.04)^{\circ}$ per period \cite{Abuter}, which is in excellent agreement with the prediction by General Relativity (GR) $\Delta \phi_{\rm GR}=0.2^{\circ}$. Based on this constraint, we can rule out the range $5.4 \times 10^{-19}$ eV $\le m \le 4.8 \times 10^{-17}$ eV for $M_h=(1.0-1.4) \times 10^{12}M_{\odot}$. The effect of the uncertainty in $M_h$ has a very little effect on the ruled out range only.

Generally speaking, the reversal in precession occurs when the orbital semimajor axis is close to the ULDM core radius. The motion of the star would be affected significantly when it is crossing the `boundary' of the ULDM core. It is because the ULDM density changes abruptly near the soliton core region. If $m$ is too small, the orbit is entirely inside the soliton core and the core density may be too small to affect the orbital precession. If this is the case, the precession angle would approach the GR prediction. If $m$ is too large, the core radius is too small so that the whole orbit is outside the core region. As the ULDM density drops significantly outside the core radius, the overall effect on precession is similar to adding a small point mass inside the orbit. In other words, different positions and sizes of the stellar orbits might generate different changes or behaviours in the precession angle.

Although the S2 star is the only one in the S-Star cluster which has been confirmed in the orbital angle precession, we expect that some other stars would soon complete one orbital period so that we can determine the precession angle more precisely. In the S-Star cluster, we have chosen 2 more representative stars (S12 and S4716) for analyses. The S12 star has a large semimajor axis and a high orbital eccentricity. Although it has a relatively long period ($P \approx 59$ years), it has been monitored for almost 3 decades already \cite{Gillessen}. We need not wait for a very long time to determine the precession angle. Also, it has an appropriate orbital inclination angle so that less uncertainty would be resulted in orbital determination. The S4716 star is a recently discovered star which has only a very short orbital period ($P \approx 4$ year) \cite{Peibker}. It has a very small semimajor axis so that it can help constrain the large regime of $m$ in the ULDM model. The orbital parameters of these 3 stars are shown in Table 1.

In Fig.~1, we plot the precession angle against $m$ for the 2 stars for comparison. We can see that a large reverse in the precession angle is resulted for the S12 star. The maximum retrograde precession is $0.9^{\circ}-1.1^{\circ}$ per period (negative sign represents the retrograde precession) when $m \approx 2\times 10^{-18}$ eV. The retrograde precession would occur if $2.8 \times 10^{-19}$ eV $<m<1.8\times 10^{-17}$ eV. The precession angle predicted by GR is $0.088^{\circ}$ per period (prograde). For the S4716 star, there is no retrograde precession, except $M_h$ being close to $1.4\times 10^{12}M_{\odot}$. The dark matter soliton core can significantly suppress the precession angle from the GR predicted value $\Delta \phi_{\rm GR}=0.26^{\circ}$ to nearly no precession ($\Delta \phi \approx 0^{\circ}$) when $m \approx 1.1\times 10^{-17}$ eV. We expect that the precession angle of the S4716 star can be precisely determined after 4 years. Therefore, we can critically examine the ULDM model for $m \sim 10^{-17}$ eV in the near future. In Fig.~2, we show the predicted orbital precession of the S2, S12, and S4716 stars after 20 periods for illustrations and the core radii of the ULDM halo for $m=10^{-18}$ eV ($\approx 0.4$ arcsec) and $m=10^{-17}$ eV ($\approx 0.04$ arcsec) for comparison.

\section{Discussion}
We have simulated the predicted effects on the stellar orbital precession angle due to the existence of the dark matter soliton core. These effects have not been discussed or investigated previously. For a certain range of ULDM mass $m$, some particular stellar orbits might undergo a reversed orbital precession. Therefore, if a dark matter soliton core exists and $m$ is within the range $10^{-19}-10^{-17}$ eV, it is possible to see the retrograde precession of some stars near the supermassive black hole. Even if we cannot observe any retrograde precession, we can constrain the value of $m$ by using the observed precession angles of different stars, such as the constraints of $m$ obtained in our study using the S2 data. Using the recent observational data of the S2 star \cite{Abuter}, we can rule out $5.4\times 10^{-19}$ eV $\le m \le 4.8\times 10^{-17}$ eV. Therefore, it is a good method to test the ULDM model if we have the information of the orbital precession for different stars inside the S-Star cluster.

Generally speaking, using the data of the stars with larger semimajor axis can give better constraints for the smaller regime of $m$. The effect of the potential retrograde precession would also be more obvious. However, the orbital periods for those stars would be very large ($>60$ years) so that we need to wait for a very long time to see the effect of orbital precession. Moreover, the effects due to the neutron stars and white dwarf mass distributions on those stellar orbital precession would also be larger. This would enlarge the systematic uncertainties in the analysis.

In fact, some studies have already use this method to examine the alleged extended mass distribution \cite{Heibel,Rubilar,Nucita} and dark matter distribution \cite{Zakharov,Dokuchaev,Dokuchaev2,Arguelles,Jovanovic,Chan2} around the Sgr A*. Different mass distributions would somewhat decrease the precession angle by a small amount only \cite{Jovanovic,Chan2}, except for some specific dark matter models \cite{Arguelles}. The overall effect on the precession depends on many factors, such as the dark matter mass, central density, scale radius, and the functional form of the distribution. However, the effect on the stellar orbital precession due to the dark matter soliton core is quite different. The sharp decrease on the density profile outside the soliton core can produce a featured variation in the precession angle. For the wave ULDM scenario, if the total halo mass $M_h$ is known, the effect on precession depends on a single parameter, the ULDM mass $m$, only. For instance, if $m=4\times 10^{-19}$ eV, we may be able to observe the star S12 undergoing retrograde precession while the S2 and S4716 stars undergoing prograde precession. In the coming decade, some stars will finish their first periods since our first observations on them (e.g. S38, S55, S62, S4711, S4714 and S4716) \cite{Gillessen,Becerra,Peibker,Peibker2,Peibker3}. High-quality observations of these stars can help provide rich information on their orbital properties and determine the precession angles. These data can critically examine the prediction of General Relativity and verify the ULDM model. Besides, the precession data can also help verify or falsify the other dark matter models (e.g. WIMP model) or constrain the extended mass distribution near the supermassive black hole.

\section{Acknowledgments}
We thank the anonymous referee for useful constructive feedback and comments. The work described in this paper was partially supported by the Seed Funding Grant (RG 68/2020-2021R) and the Dean's Research Fund (activity code: 04628) from The Education University of Hong Kong.

\begin{table}
 \vskip 3mm
\caption{Parameters of the stellar orbits \cite{Gillessen,Peibker}}
\begin{tabular}{ |l|c|c|c| }
 \hline\hline
  Star& Semimajor axis $a$ (AU) & Eccentricity $e$ & $\Delta \phi_{\rm GR}$ (deg) \\
 \hline
  S2 & 1002 & 0.886 & 0.21 \\
  S12 & 2390 & 0.888 & 0.088 \\
  S4716 & 398 & 0.756 & 0.26 \\
 \hline\hline
\end{tabular}
 \vskip 3mm
\end{table}

\begin{figure}
\vskip 3mm
\includegraphics[width=70mm]{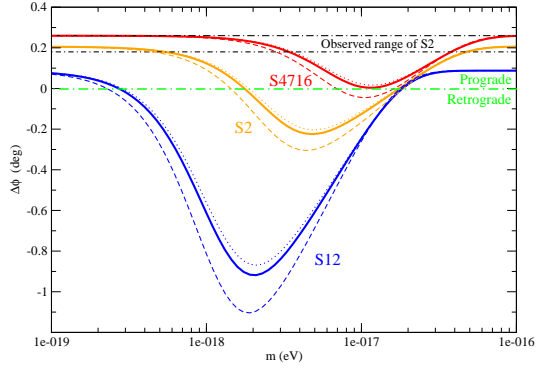}
\caption{The coloured solid lines, dotted lines, and dashed lines represent the precession angles $\Delta \phi$ against ULDM mass $m$ for $M_h=1.08\times 10^{12}M_{\odot}$, $M_h=1.0 \times 10^{12}M_{\odot}$ and $M_h=1.4 \times 10^{12}M_{\odot}$ respectively for different stars (Orange: S2; Blue: S12; Red: S4716). The region bounded by the black dash-dotted lines indicates the allowed $1 \sigma$ range observed for the S2 star \cite{Abuter}.}
\label{Fig1}
\vskip 3mm
\end{figure}

\begin{figure}
\vskip 3mm
\includegraphics[width=70mm]{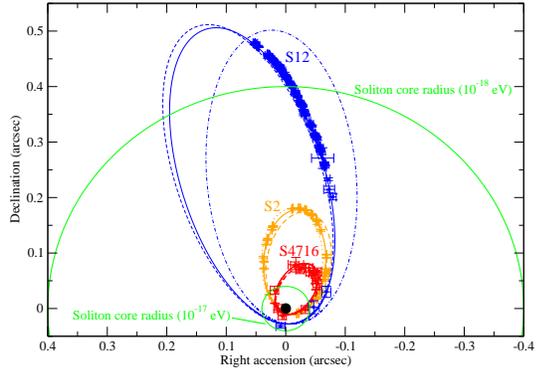}
\caption{The orange, blue and red solid lines represent the current predicted orbits of the S2, S12 and S4716 stars respectively. The corresponding dotted lines, dot-dashed lines and dashed lines indicate the precessed orbits after 20 periods for $m=10^{-17}$ eV, $m=10^{-18}$ eV and $m=10^{-19}$ eV respectively. The green solid lines represent the dark matter soliton core radii for $m=10^{-17}$ eV and $m=10^{-18}$ eV. The observed data with error bars for the S2 \cite{Do}, S12 \cite{Gillessen} and S4716 \cite{Peibker} are shown for reference. The position of the supermassive black hole (Sgr A*) is indicated by the black dot. Here, $M_h=1.08\times 10^{12}M_{\odot}$ is assumed.}
\label{Fig2}
\vskip 3mm
\end{figure}

\end{document}